\documentclass[aps,prl,twocolumn,floatfix,english,superscriptaddress,showpacs,10pt]{revtex4-2}
\usepackage{xcolor}
\usepackage{subfigure}
\usepackage{amsmath,bm}
\usepackage{mathtools}
\usepackage{commath}
\usepackage{amssymb}
\usepackage{graphicx}
\usepackage{babel}
\usepackage{cases}
\usepackage[colorlinks=true, citecolor=blue, anchorcolor=green]{hyperref}
\hypersetup{linkcolor=red}
\usepackage{natbib}
\usepackage{floatrow}
\usepackage{dblfloatfix}
\usepackage{braket}
\usepackage{lipsum}
\usepackage{float}

\makeatletter

\newcommand{\Rmnum}[2]{\expandafter\@slowromancap\romannumeral #1@}

\makeatother

\begin{document}
\title{Efficient Spin Seebeck and Spin Nernst Effects of Magnons in Altermagnets}
        \affiliation{National Laboratory of Solid State Microstructures, School of Physics, Collaborative Innovation Center of Advanced Microstructures, Nanjing University, Nanjing 210093, China} 
        \affiliation{School of Physics, Huazhong University of Science and Technology, Wuhan 430074, China}
        \affiliation {Ningbo Institute of Materials Technology and Engineering, Chinese Academy of Sciences, Ningbo 315201, China} 
        \affiliation{Yongjiang Laboratory, Ningbo 315202, China}
        
        \author{Qirui Cui}  
        \affiliation{National Laboratory of Solid State Microstructures, School of Physics, Collaborative Innovation Center of Advanced Microstructures, Nanjing University, Nanjing 210093, China}
        \affiliation {Ningbo Institute of Materials Technology and Engineering, Chinese Academy of Sciences, Ningbo 315201, China} 
        \affiliation{Yongjiang Laboratory, Ningbo 315202, China}
        
        \author{Bowen Zeng}
        \affiliation{School of Physics, Huazhong University of Science and Technology, Wuhan 430074, China}
        
        \author{Tao Yu}       
        \email{taoyuphy@hust.edu.cn}
        \affiliation{School of Physics, Huazhong University of Science and Technology, Wuhan 430074, China}
        
        \author{Hongxin Yang}
        \email{hongxin.yang@nju.edu.cn} 
        \address{National Laboratory of Solid State Microstructures, School of Physics, Collaborative Innovation Center of Advanced Microstructures, Nanjing University, Nanjing 210093, China}
        
        \author{Ping Cui}
        \affiliation {Ningbo Institute of Materials Technology and Engineering, Chinese Academy of Sciences, Ningbo 315201, China} 
        \affiliation{Yongjiang Laboratory, Ningbo 315202, China}
           
\begin{abstract}
We report two non-degenerate magnon modes with opposite spins or chiralities in collinearly antiferromagnetic insulators driven by symmetry-governed anisotropic exchange couplings. The consequent giant spin splitting contributes to spin Seebeck and spin Nernst effects generating longitudinal and transverse spin currents when the temperature gradient applies along and away from the main crystal axis, without requiring any external magnetic field and spin-orbit coupling. Based on first-principle calculations,
we predict feasible material candidates holding robust altermagnetic spin configurations and room-temperature structural stability to efficiently transport spin. \textcolor{red}{The spin Seebeck conductivity is comparable to the records of antiferromagnets that require the magnetic field, and the spin Nernst conductivity is two orders in magnitude larger than that in antiferromagnetic monolayers that need Berry curvature.} 
\end{abstract}

\maketitle

\emph{Introduction}.---Magnons in magnetic insulators favor the spin transport with low energy consumption \cite{am1, am2, am3, am4, am5, am6, am7, am8, am9, am10, sam1}. Although the chirality of magnon is fixed in ferromagnets, symmetry allows two degenerate magnon modes carrying opposite spins in collinear antiferromagnets with easy-axis anisotropy \cite{am11, am12, am13, am14, am15, am16}. With no net magnetization, vanished static stray field, and terahertz response rate, antiferromagnets exhibit unique advantages for robust and ultrafast spin-based nanoscale applications, among which the spin Seebeck effect (SSE) and spin Nernst effect (SNE) are, respectively, effective approaches to longitudinally and transversely transport spins with respect to the temperature gradient~\cite{am12, am14, am16, am17, am18, am19, am20, am21, am22, am23, vs1, am24, am25, am26, am27, am28, am29, am30, am31, am32, am33}. In collinear antiferromagnets, the SSE often requires a strong magnetic field to break the spin degeneracy \cite{am11, am12, am24}, and more stringently, the SNE exploits the Berry curvature that needs the Dzyaloshinskii-Moriya interaction (DMI) \cite{am14, am25, am34, am35} or magnetoelastic coupling \cite{am31, am32}, i.e., the transverse spin current is normally proportional to the spin-orbit coupling (SOC) strength. 
\textcolor{red}{Theoretical effort made to realize spin transport free of external magnetic field and SOC is based on the noncollinear antiferromagnetic insulator of Kagome lattice \cite{rp1}. Nevertheless, a large magnetic field appears to be necessary to ensure the identical spin configurations between Kagome layers. Also, the symmetry in such Kagome lattice still allows for the nearest-neighboring DMI, which is used to modulate the spin transport ~\cite{rp2}. Therefore, it is fair to say the spin transport via magnons, which can be $entirely$ free of the magnetic field and Berry curvature, in antiferromagnets remains wanting.}

Metallic altermagnet is a special type of collinear antiferromagnet without net magnetization that was recently reported to hold electronic bands with spin splitting in the reciprocal space, even in the absence of any relativistic SOC,  due to breaking of the combined space inversion, time reversal, and translation symmetries \cite{rp3, rp4, am36, am37, am38, am381, am382}. By applying the external electrical field, the motion of the Fermi surface with spin splitting could result in the strong transverse spin current, similar to the spin Hall effect, which was experimentally demonstrated based on the altermagnetic metal $\rm RuO_2$ \cite{sam2, sam3, sam4}. Interestingly, antiferromagnetic magnons possessing the non-degenerate chirality have recently been predicted in $\rm RuO_2$ as well \cite{sam5}. However, the spin transport properties dominated by such magnons are still unclear, for which the insulators can avoid the complications from electrons in metals.

\begin{figure}[b]
\includegraphics[width=1\linewidth]
{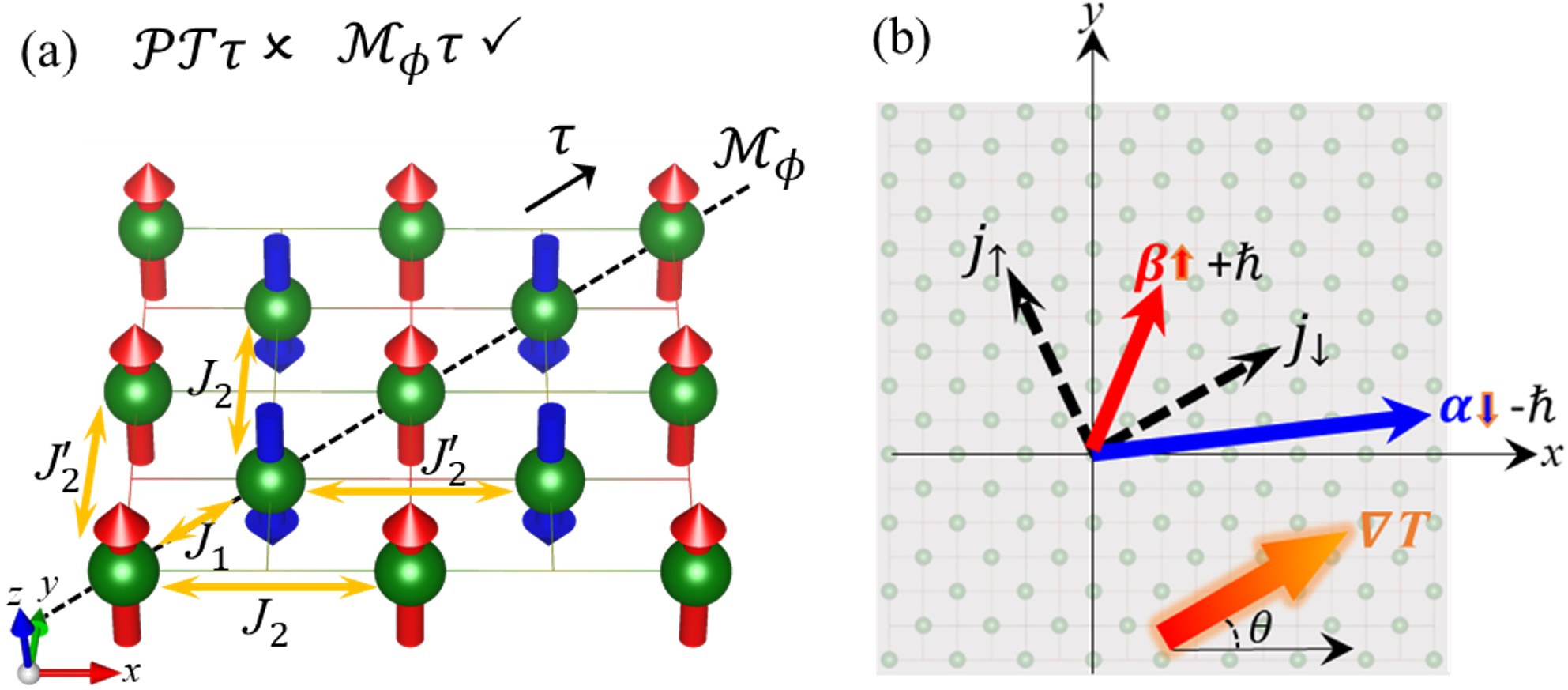}
\caption{(a) The schematic of two-dimensional antiferromagnetic spin-lattice, without $\mathcal{PT} \tau$ symmetry but holding $\mathcal{M}_{\phi} \tau$ symmetry. The red and blue arrows represent spin vectors in the out-of-plane orientation. (b) A thermal gradient away from the main axis drives longitudinal and transverse spin currents, contributed by different magnon modes with opposite spins.}
\label{Fig1}
\end{figure}

In this Letter, we report two spin non-degenerate $\alpha$ and $\beta$ magnon modes allowed by symmetry in collinear antiferromagnets, driven by the anisotropic spin exchange couplings [Fig.~\ref{Fig1}(a)]. This leads to the anisotropic spin-momentum locking for magnons favoring efficient SSE and SNE, which are free of external magnetic fields or Berry curvature. Specifically, nonequivalent magnon currents carried by $\alpha$ and $\beta$ modes with opposite spins are stimulated by the temperature gradient but flow along different directions, which makes their combination produce longitudinal and transverse spin currents [Fig.~\ref{Fig1}(b)]. 
Based on first-principle calculations, we propose potential material candidates, e.g. $\rm Cr_2Te(Se)_2O$ monolayer, for observing these phenomena. 
Our findings offer the missing piece of spin transport in altermagnets by opening a novel way towards producing spin current via antiferromagnetic magnons and providing an interesting platform for exploring spin caloritronic effects.

\emph{Model}.---We construct a two-dimensional antiferromagnetic square lattice with non-relativistic spin splitting by symmetry analysis [Fig.~\ref{Fig1}(a)]. The lacking of combined space inversion $\mathcal{P}$, time reversal $\mathcal{T}$ and half-unit cell translation $\tau$ symmetries ($\mathcal{PT}\tau$) induces the spin-splitting for electronic eigenstates, while the preservation of combined mirror
$\mathcal{M}_{\phi}$ and $\tau$ symmetries ($\mathcal{M}_{\phi}\tau$) guarantees eigenstates holding opposite spins at $\bf k$ and $\mathcal{M}_{\phi} \bf k$, thus tuning the conventional antiferromagnet to altermagnet \cite{am37, am39, am40}. Notably, from the structural aspect, $\mathcal{P}\mathcal{T}\tau$ symmetry breaking requires the nearest-neighboring spin sites in each sublattice being connected anisotropically in the $\hat{\bf x}$-  and $\hat {\bf y}$-directions. Thereby, the bridged electronic states give rise to anisotropic superexchange coupling based on Goodenough-Kanamori-Anderson (GKA) theory~\cite{am41, am42, am43}. Further, $\mathcal{M}_{\phi}\tau$ symmetry requires spin-connected types of sublattices ``a'' and ``b'' being exactly swapped in $\hat{\bf x}$  and $\hat {\bf y}$, and N\'{e}el magnetization paralleling with $\mathcal{M}_{\phi}$, which results in the switched exchange coupling for two sublattices. The symmetry-governed spin Hamiltonian  
\begin{flalign} 
\hspace{-2mm}
\hat{H} = & J_1 \! \sum_{\langle i, j\rangle} \hat{\bf {S}}_{a i} \! \cdot \! \hat{\bf S}_{b j}\nonumber\\
+ & J_2 \! \sum_{\langle i_x, j_x\rangle} \! \hat{\bf{S}}_{a i}\! \cdot \! \hat{\bf{S}}_{a j} \!
+ \! J_2' \! \sum_{\langle i_y, j_y\rangle} \! \hat {\bf{S}}_{a i} \! \cdot \! \hat {\bf{S}}_{a j} \! + \! K \sum_i \! \left(\hat{S}_{a i}^z\right)^2 \nonumber\\
+ &  J_2' \!  \sum_{\langle i_x, j_x\rangle} \! \hat {\bf{S}}_{b i} \! \cdot \! \hat {\bf{S}}_{b j} \! + \! J_2 \sum_{\langle i_y, j_y\rangle} \! \hat {\bf{S}}_{b i} \! \cdot \! \hat{\bf{S}}_{b j} \! + \! K \sum_i \! \left(\hat{S}_{b i}^z\right)^2
\label{e1} 
\end{flalign}
describes explicitly these magnetic properties,
where $\hat{\bf S}_a$ and $\hat{\bf S}_b$ represent the spin operators of the same magnitude. $J_1 > 0$ favors the antiferromagnetic inter-sublattice exchange coupling, $J_2 (J_2^{\prime})<0$ favors the ferromagnetic intra-sublattice exchange coupling, and single-ion magnetic anisotropy $\emph{K}<0$ favors the out-of-plane magnetization. $\langle i, j \rangle$, $\langle i_x, j_x \rangle$ and $\langle i_y ,j_y \rangle$ indicates the nearest-neighboring coupling along the diagonal, $\hat{\bf x}$ and $\hat{\bf y}$, respectively. 

Using the Holstein-Primakoff transformation~\cite{am44}, the Hamiltonian in terms of bosonic operators $\hat{\psi}_{\bf k}=\left(\hat{a}_{\bf k}, \hat{b}_{\bf k}^{\dagger} \right)^T$ in the momentum space reads $\hat{H} = \sum_{\bf k} \hat{\psi}_{\bf k}^{\dagger} \hat{H}_{\bf k} \hat{\psi}_{\bf k}$, where 
\begin{align}
\hat{H}_{\bf k} = S \left(\begin{array}{cc} J^* + 2 J_2 \gamma_1 + 2 J_2^{\prime} \gamma_2 & 4 J_1 \gamma_3 \\ 4 J_1 \gamma_3 & J^* + 2 J_2^{\prime} \gamma_1 + 2 J_2 \gamma_2 \end{array} \right).
\end{align}
Here the on-site energy $J^*=4 J_1-2 J_2 - 2 J_2^{\prime}-2K$, and the structure factors $\gamma_1=\cos(k_xa)$, $\gamma_2=\cos(k_ya)$, $\gamma_3=\cos(k_xa/2)\cos(k_ya/2)$. This Hamiltonian is diagonalized by the Bogoliubov transformation $\hat{\psi}_{\bf k} = T_{\bf k} \hat{\Psi}_{\bf k}$ in terms of the magnon operators of $\alpha$ and $\beta$ modes $\hat{\Psi}_{\bf{k}}=\left( \hat{\alpha}_{\bf k}, \hat{\beta}_{\bf k}^{\dagger} \right)^T$: 
\begin{align} 
\hat{H}=\sum_{\bf k} \hat{\Psi}_{\bf{k}}^{\dagger} \left(\begin{array}{cc} \hbar\omega_{\bf k}^\alpha & 0 \\ 0 & \hbar\omega_{\bf k}^\beta \end{array}\right) \hat{\Psi}_{\bf{k}},
\end{align}
where the eigenfrequencies of $\alpha$ and $\beta$ modes
 \begin{align}
&\omega_{\bf k}^\alpha=\frac{S}{\hbar} \left[(J_2-J_2^{\prime})\left(\gamma_1-\gamma_2\right)+\frac{4J_1\gamma_3 \Delta}{\gamma_e}\right], \nonumber\\
&\omega_{\bf k}^\beta=\frac{S}{\hbar} \left[(J_2^{\prime}-J_2)\left(\gamma_1-\gamma_2\right)+\frac{4J_1\gamma_3 \Delta}{\gamma_e}\right].
\end{align}
Here $\Delta=\sqrt{1-{\gamma_e}^2}$ and effective form factor $\gamma_e= {4J_1\gamma_3}/[{4J_1-2K-(J_2+J_2^{\prime})(2-\gamma_1-\gamma_2)}]$. Importantly, $\alpha$ and $\beta$ modes are non-degenerate in the reciprocal space, except for $\mathrm M \! \leftrightarrow \! \Gamma$ lines at which $\left(\gamma_1-\gamma_2\right)=0$. The corresponding frequency splitting
\begin{align}
\Delta \omega = \frac{2S}{\hbar}| \left( J_2-J_2^{\prime} \right) \left( \gamma_1-\gamma_2 \right)|. 
\end{align}

$J_2\!\neq\!J_2^{\prime}$ in each sublattice is induced by the $\mathcal{P}\mathcal{T}\tau$ symmetry breaking, while $J_2$-$J_2^{\prime}$ swapping is protected by the $\mathcal{M}_{\phi}\tau$ symmetry, which render such mode splitting. Besides spectrum splitting, $\alpha$ and $\beta$ modes exhibit opposite chiralities and spins. Via Bogoliubov transformation, the creation operators of $\alpha$ and $\beta$ modes $\hat{\alpha}_{\bf k}^{\dagger} = u_{\bf k} \hat{a}_{\bf k}^{\dagger} - v_{\bf k} \hat{b}_{\bf k}$ and $\hat{\beta}_{\bf k}^{\dagger} = u_{\bf k} \hat{b}_{\bf k}^{\dagger} - v_{\bf k} \hat{a}_{\bf k}$.
$\hat{S}_i^{+}$ and $\hat{S}_i^{-}$ generate opposite spin precessions at site $i$, which are switched between sublattices ``a'' and ``b''. Accordingly, the dynamics of  $\hat{\alpha}_{\bf k}^{\dagger}$ ($\hat{\beta}_{\bf k}^{\dagger}$) is governed by right-handed (left-handed) circular polarization in ``a'' and ``b''.
For $\alpha$ and $\beta$ magnon in antiferromagnets with out-of-plane magnetization, the  $\hat{\bf z}$-component spin angular momentum with unit of $\hbar$ is opposite since  $\langle0|\hat{\alpha}_k S^z \hat{\alpha}_k^{\dagger}|0\rangle=-1$ and $\langle 0|\hat{\beta}_k S^z \hat{\beta}_k^{\dagger}|0\rangle=1$. Our model thus elucidates the basic properties of altermagnetic magnons, including spectrum splitting, chirality, and chirality-locked spin.

The out-of-plane magnetization also opens up a magnon excitation gap of $\left(2S/\hbar\right)\sqrt{K^2-4J_1K}$ at $\Gamma$ point, which is crucial for stabilizing two-dimensional magnetism with long-range order by overcoming restrictions of Mermin-Wanger theorem \cite{am45, am46, am47}. We detail the construction of magnon Hamiltonian in the Supplemental Materials (SM)~\cite{am48}.

\begin{figure}[t]
\includegraphics[width=1\linewidth]{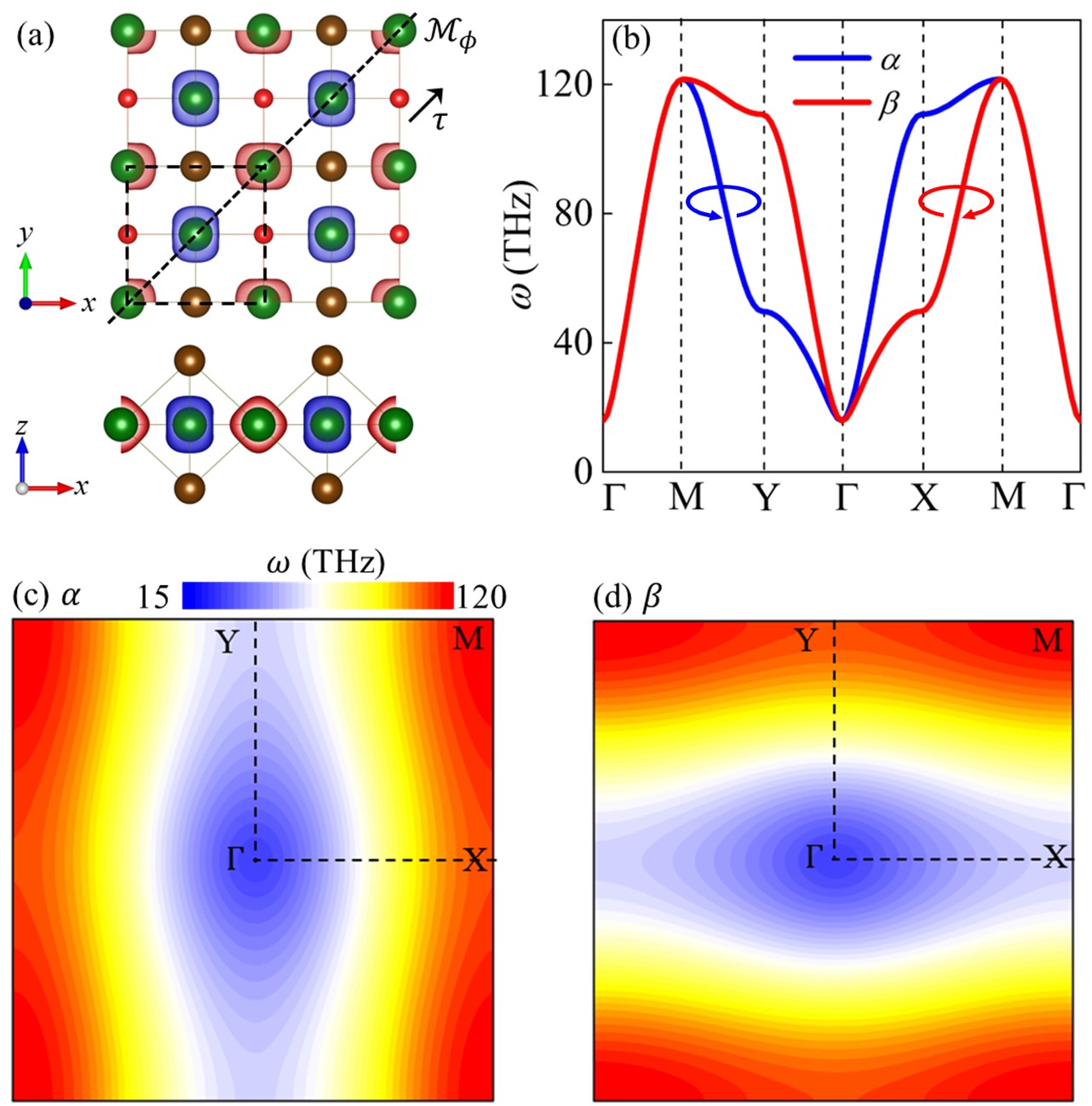}
\caption{(a) The top and side views of the crystal structure of CTO. The dashed square indicates a unit cell with the optimized lattice constant $a=4.12$ \AA. The bonding angle of Cr–Te–Cr and Cr-O-Cr equal $\rm 94^{\circ}$ and $\rm 180^{\circ}$, respectively. The blue and red zones located around Cr elucidate the spin density distribution in the real space. (b) Magnon band structure with spin splitting. (c) and (d) Magnon spectrum of $\alpha$ and $\beta$ modes in the whole Brillouin zone.} 
\label{Fig2}
\end{figure}

\emph{Material candidates}.---Next, we show chromium oxytelluride, $\mathrm{Cr_2Te_2O}$ (CTO) monolayer [see crystal structure in Fig.~\ref{Fig2}(a)] as a promising insulator for transporting altermagnetic magnons, referring to SM~\cite{am48} for details of the first-principle calculations. CTO possesses the $P4/mmm$ space group, and the existence of bridging Te and O between Cr breaks $\mathcal{PT}\tau$ symmetry [Fig.~\ref{Fig2}(a)]. The octahedral crystal filed around Cr approximately splits 3$d$ orbitals into the lower energy set $t_{2g}$ and higher energy set $e_g$. Accordingly, each Cr carries a net magnetic moment around $3.29\mu_B$, which is consistent with the spin number $S=3/2$ of $\mathrm{Cr^{3+}}$. By using energy mapping methods, we calculate $J_1=3.90$, $J_2 =-7.90$, $J_2^{\prime}=-1.21$ and $K=-0.73$ meV, exactly favoring the spin configuration illustrated in Fig.~\ref{Fig1}(a). Since the bonding angle of Cr–Te–Cr is close to $\rm 90^{\circ}$, according to GKA rules, the spin superexchange mediated by intervening Te prefers to be ferromagnetic, while the bonding angle of Cr–O–Cr reaching $\rm 180^{\circ}$ favors the antiferromagnetic superexchange, which is responsible for the difference between $J_2$ and $J_2^{\prime}$. Monte Carlo simulations based on the explicit spin Hamiltonian [Eq.~(\ref{e1})] determine the N\'{e}el temperature reaching 205 K~\cite{am48}.

Figure~\ref{Fig2}(b) plots the magnon spectrum of CTO. Except for $\mathrm M \! \leftrightarrow \! \Gamma$ lines, the frequency of $\alpha$ and $\beta$ modes splits significantly with the largest splitting of up to 63.39~THz appearing at X and Y points. Magnonic anisotropy is more clearly illustrated in Fig.~\ref{Fig2}(c) and (d). For the $\alpha$ mode, the more rapid frequency variation along  $\hat{\bf x}$ than $\hat{\bf y}$ implies the much larger group velocity $v_x$ than $v_y$ \cite{am48}, leading to thermal orientation-dependent magnonic transport behavior, see below. The situation is similar for the $\beta$ mode but the rapid frequency variation emerges along $\hat{\bf y}$. Since the two modes carry opposite spin, the above results reveal that the altermagnetic magnon nondegeneracy is characterized by anisotropic spin-momentum locking. 
Crucially, CTO exhibits insulating features as confirmed by the indirect electronic gap of 0.30 eV \cite{am48}, indicating negligible thermal electron excitation at low temperatures. Without electronic contribution, the energy dissipation via joule heating can be avoided, and more importantly, the observation and application of thermal/spin current contributed by pristine magnons become possible.

We also find alternate 
insulating chromium oxyselenide, $\mathrm{Cr_2Se_2O}$ monolayer with magnetic parameters $J_1=12.48$, $J_2=-5.80$, $J_2^{\prime}=6.69$ and $K=-0.30$ meV holds magnon modes with anisotropic spin-momentum locking \cite{am48}. \textcolor{red}{Notably, the DMI is forbidden for proposed candidates based on the Moriya symmetry rules~\cite{mor1}}.  The phonon dispersion without any imaginary frequency in the entire Brillouin zone and $ab~initio$ molecular dynamic simulations showing the stable internal energy evolution at 500~K \cite{am48} demonstrates their dynamical and thermal stability.

\begin{figure}[t]
\includegraphics[width=1\linewidth]{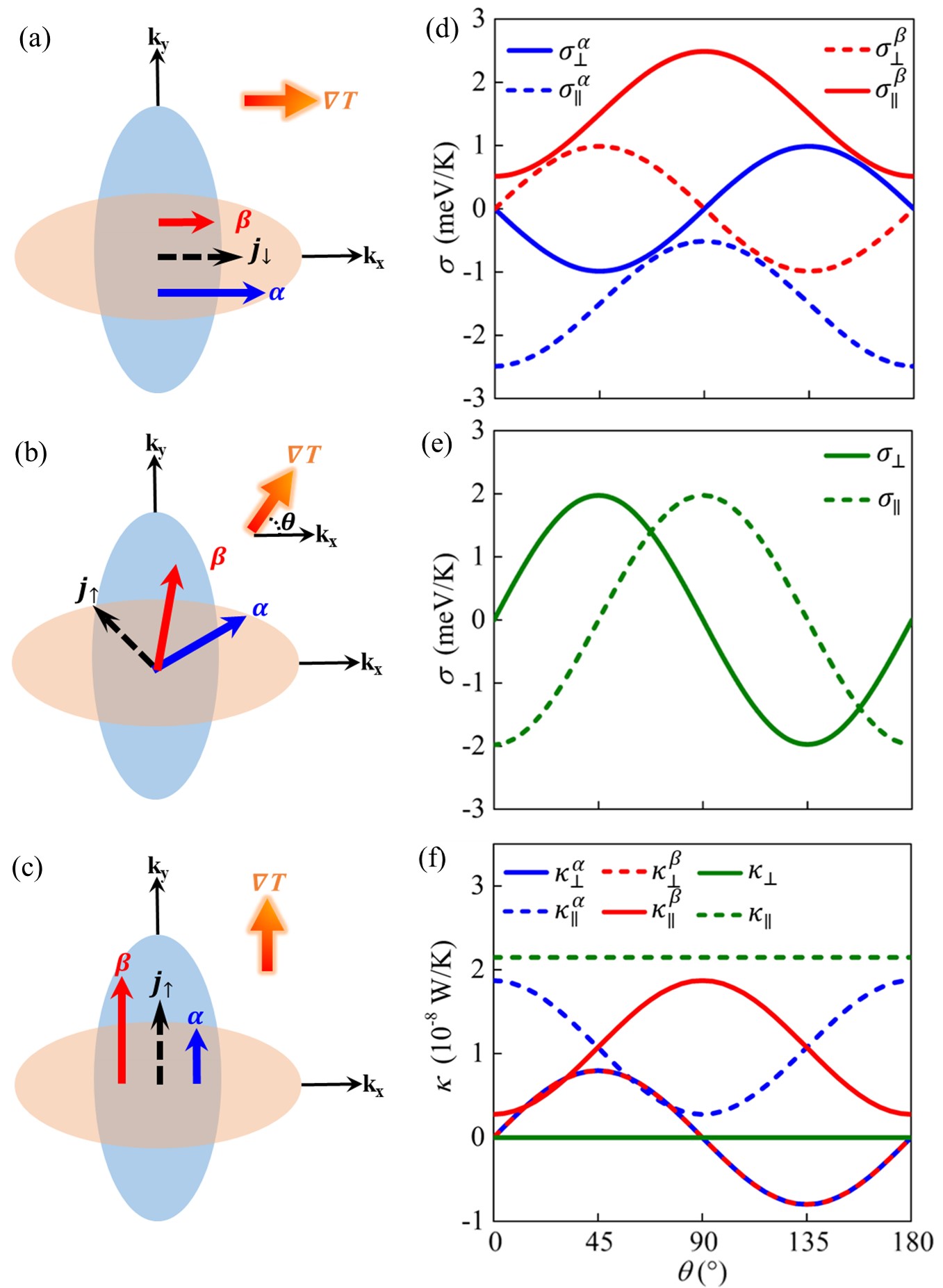}
\caption{(a)-(c) The flow of spin-polarized magnons driven by different temperature gradients. When the gradient direction is away from the main crystal axis (a) $\hat{\bf x}$ and (c) $\hat{\bf y}$, the currents carried by $\alpha$- and $\beta$-magnon are no longer parallel to its orientation, leading to the transverse spin current in (b). (d) and (e) show the calculated spin conductivity and (f) is thermal conductivity. $\tau_0=10$~ps and $T=100$~K.}
\label{Fig3}
\end{figure}

\emph{Spin Seebeck and Spin Nernst effects}.---The magnon current can be conveniently driven by the temperature or chemical potential bias. In the linear response regime, such spin current density under the temperature $T$ gradient reads 
\begin{align}
\left(\begin{array}{cc} j_x^{z} \\ j_y^{z} \end{array}\right)= \left(\begin{array}{cc}\sigma_{xx} & 0 \\ 0 & \sigma_{yy}\end{array}\right) \left(\begin{array}{cc} \left(-\partial_x T \right) \rm{cos}\theta \\ \left(-\partial_y T \right) \rm{sin}\theta \end{array}\right), \label{e8} \end{align}
where $\theta$ is the direction of the temperature gradient with respect to the $\hat{\bf x}$-direction [Fig.~\ref{Fig3}(b)]. Based on the Kubo formula~\cite{rp1,am50,am51,am52,yu1}, the magnon thermal conductivity tensor $\sigma_{mn} = \sigma_{mn}^{\alpha} + \sigma_{mn}^{\beta}$ is contributed by the $\alpha$ and $\beta$ mode separately
with~\cite{am48}
\begin{align}
\sigma_{mn}^{\alpha} \! = \! -  \frac{\tau_0}{A k_{B} T^2} \! \sum_{\bf k} \! \left(\hat{V}_m \right)_{\alpha \alpha} \! \left(\hat{V}_n \right)_{\alpha \alpha} \! \omega_{\bf k}^{\alpha} \! \frac{e^{\hbar \omega_{\bf k}^{\alpha}  /k_{B}T}}{\left( \! e^{\hbar \omega_{\bf k}^{\alpha} /  k_{B}T}  -  1 \! \right)^2}, \nonumber \\
\sigma_{mn}^{\beta} \! = \! \frac{\tau_0}{A k_{B} T^2} \! \sum_{\bf k} \! \left(\hat{V}_m \right)_{\beta \beta} \! \left(\hat{V}_n \right)_{\beta \beta} \! \omega_{\bf k}^{\beta} \! \frac{e^{-\hbar \omega_{\bf k}^{\beta}  /  k_{B}T}}{\left( \! e^{-\hbar \omega_{\bf k}^{\beta} /  k_{B}  T} - 1 \! \right)^2}, \nonumber
\end{align}
where $n$ is along the temperature gradient, $m$ represents the direction of driven spin current, $A$ is the sample area, $\tau_0$ is magnon lifetime, and $\hat{V}_m\!=\!T_{\bf {k}}^{\dagger} (\partial_{k_m}\hat{H}_{\bf k})  T_{\bf {k}}$. The same results are obtained by the semi-classical Boltzmann transport theory with relaxation-time approximation~\cite{am48}. Notably, the spin Berry curvature $\Omega(\bf k)$ vanishes due to the coexistence of $\mathcal{P}$ and $\mathcal{T}c_x$ symmetries in our altermagnet model~\cite{am14}, where $c_x$ represents $180^{\circ}$ rotation operation around the $\hat{\bf x}$-axis. Taking  $\tau_{0}=10$~ps \cite{am54, am55, am56} and $T=100$~K, we find $\sigma_{xx}^{\alpha} = - \sigma_{yy}^{\beta} = -2.49$ and $\sigma_{yy}^{\alpha} = - \sigma_{xx}^{\beta} = -0.51$~meV/K in CTO. 

By projecting to the gradient direction, we find the longitudinal and transverse spin currents 
 \begin{align}
\left(\begin{array}{cc} j_\parallel^{z} \\ j_\perp^{z} \end{array}\right)= \left(\begin{array}{cc}\sigma_{\parallel} & 0 \\ 0 & \sigma_{\perp}\end{array}\right) \left(\begin{array}{cc} -\partial_n T \\ -\partial_n T \end{array}\right),
\label{e11} 
\end{align}
where $\sigma_{\parallel} = \sigma_{\parallel}^{\alpha} + \sigma_{\parallel}^{\beta}$ and $\sigma_{\perp} = \sigma_{\perp}^{\beta} - \sigma_{\perp}^{\alpha}$ with $\sigma_{\parallel}^{\alpha} \! = \! \sigma_{xx}^{\alpha} {\rm cos}^2\theta + \sigma_{yy}^{\alpha} {\rm sin}^2\theta,~ \sigma_{\parallel}^{\beta} \! = \! \sigma_{xx}^{\beta} {\rm cos}^2\theta + \sigma_{yy}^{\beta} {\rm sin}^2\theta$, $\sigma_{\perp}^{\alpha} \! = \!(\sigma_{xx}^{\alpha}-\sigma_{yy}^{\alpha}) {\rm cos}\theta {\rm sin}\theta$, and $~\sigma_{\perp}^{\beta} \! = \! (\sigma_{yy}^{\beta}-\sigma_{xx}^{\beta}) {\rm cos}\theta {\rm sin}\theta.$ When the temperature gradient is along the main crystal axis $\hat{\bf x}$ $(\theta=0^{\circ})$, only the longitudinal spin current is generated [Fig.~\ref{Fig3}(a) and (d)]. The nondegeneracy of magnon modes leads to $| \sigma_{xx}^{\alpha} | \neq | \sigma_{xx}^{\beta} |$ and thus generates the SSE with a spin-down current $\sigma_{\parallel}=-1.98$ meV/K [Fig.~\ref{Fig3}(e)]. The external field generating magnon imbalance is not a prerequisite for realizing SSE in altermagnets, however. 
When the temperature gradient deviates from $\hat{\bf x}$ $(0^{\circ} \textless \theta \textless 90^{\circ})$, the currents carried by $\alpha$ and $\beta$ modes are not parallel to temperature gradient because of the anisotropy  $\sigma_{xx}^{\alpha(\beta)} \neq  \sigma_{yy}^{\alpha(\beta)} $, and flow in different directions. $| \sigma_{xx(yy)}^{\alpha} | = | \sigma_{yy(xx)}^{\beta} |$ by $\mathcal{M}_{\phi}$ symmetry [Fig.~\ref{Fig3}(b) and (d)]. Consequently, the SNE with transverse spin-up current appears [Fig.~\ref{Fig3}(e)]. For $\theta=45^{\circ}$, the transverse current reaches its maximum with $\sigma_{\perp} = 1.98$~meV/K since the $\alpha$- and $\beta$-magnons flowing along the opposite directions has maximum magnitudes. The transverse spin current vanishes again as the temperature gradient rotates to the main crystal axis $\hat{\bf y}$ ($\theta = 90^{\circ}$), while the longitudinal spin-up current with $\sigma_{\parallel}=1.98$ meV/K emerges [Fig.~\ref{Fig3}(c) and (e)]. Different from spin current, the magnonic thermal current only flows along the temperature gradient with constant $\kappa_{\perp}=0$ and $\kappa_{\parallel}=2.15 \times 10^{-8}$~W/K [Fig.~\ref{Fig3}(f)], revealing the vanished thermal Hall effect. 

\begin{figure}[h]
\includegraphics[width=1\linewidth]{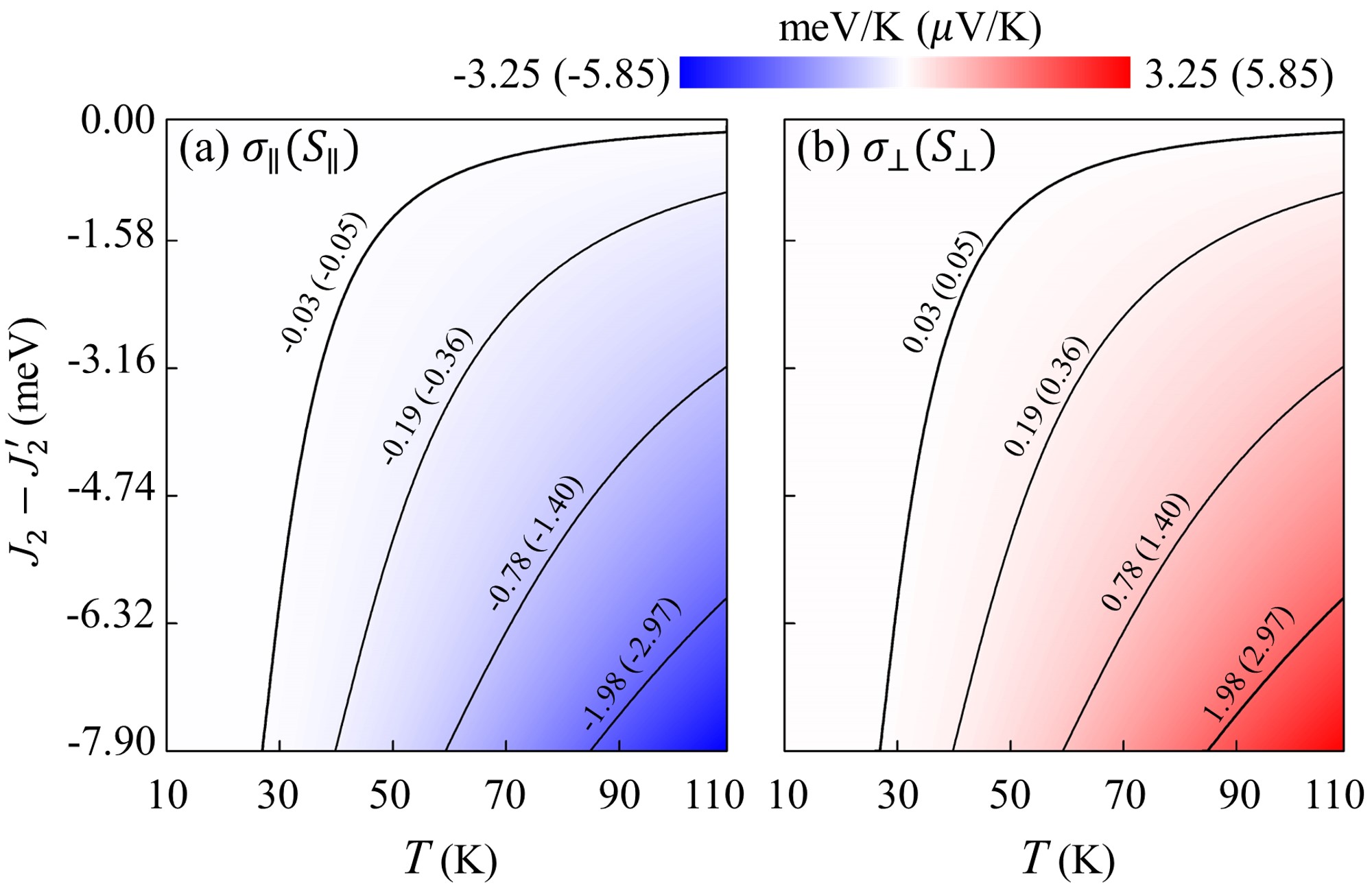}
\caption{(a) SSE and (b) SNE coefficients as a function of temperature and $J_2-J_2^{\prime}$. The temperature gradient is set along the $\hat{\bf x}$ and diagonal directions for observing SSE and SNE, respectively.}
\label{Fig4}
\end{figure}

Figure~\ref{Fig4} addresses $\sigma_{\parallel}$ and $\sigma_{\perp}$ as a function of temperature and $J_2-J_2^{\prime}$. Their magnitudes increase rapidly and monotonically with temperature or difference between $J_2$ and $J_2^{\prime}$. This is substantiated since higher temperature produces more magnon occupation by the Bose-Einstein distribution and the larger magnitude of $J_2-J_2^{\prime}$ results in strong magnonic anisotropy. \textcolor{red}{Monte Carlo simulations unveil that the normalized magnetization remains around 0.85 [Fig. S2(a)] at the temperature up to 110 K [Fig. ~\ref{Fig4}], indicating the fluctuation around the long-range magnetic order is still small. This validates the model of altermagnetic magnons based on the linear spin wave theory when the temperature range is much lower than the critical one.} Even at low temperature $T=29$~K, the SNE $\sigma_{\perp}$ reaches 0.03~meV/K that is two orders in magnitude larger than that induced by nonzero spin Berry curvature in monolayer and bilayer antiferromagnets~\cite{am14, am31, am32}.
Further, we compare the SSE coefficients in experiments by adopting the formula $\mathcal{S} = (e/\hbar d) \rho \theta_{SH} \sigma \cos(\eta)$ that provides the inverse spin Hall voltage~\cite{am33, am57, am58}, where $d$ is the interlayer distance, the electrical resistance $\rho=105$ n$\Omega \cdot$m, the spin Hall angle of Pt $\theta_{SH}=0.068$~\cite{am60}, and the bulk inclination angle $\eta=45^{\circ}$. We find that SSE is of the same order as the experimental record of antiferromagnets, where a magnetic field around 10~T is required \cite{vs1, am22, am61, am62}.

\emph{Conclusion and discussion}.---We predict the SSE and SNE in collinearly antiferromagnetic insulators with efficient heat-to-spin conversion, overcoming the limitations from Joule heat, external magnetic field, and SOC. 
\textcolor{red}{The spin Seebeck conductivity is of the same magnitude as the records in other antiferromagnets where a large magnetic field is desired for generating spin splitting, and the spin Nernst conductivity is much larger than that of antiferromagnetic few-layers induced by spin Berry curvature. Within an appropriate temperature range, both spin Seebeck and Nernst conductivities can be significantly enhanced by increasing temperature.} The strong longitudinal and transverse spin currents driven by thermal bias can be further applied to produce torque on adjacent magnets, thus providing a way for manipulating magnetization.  
The anisotropic spin-momentum locking of magnons, arising from the anisotropic exchange coupling determined by $\mathcal{PT}\tau$ symmetry breaking and $\mathcal{M}_{\phi}\tau$ symmetry preservation, \textcolor{red}{results in a distinct response of  $\alpha$- and $\beta$-magnons.} Meanwhile, the coexistence of $\mathcal{P}$ and $\mathcal{T}c_x$ symmetries excludes any role of Berry curvature on magnon propagation. Possible material candidates are also provided for realizing proposed phenomena that should widely exist in altermagnetic insulators \cite{am48} with easy-axis anisotropy. 

\begin{acknowledgments}
This work was supported by the National Key Research and Development Program of China (MOST) (Grants No. 2022YFA1405102 and 2022YFA1403601), the National Natural Science Foundation of China (Grant No. 12174405), the startup grant of Huazhong University of Science and Technology (Grants No. 3004012185 and 3004012198). This work was performed during the visit of School of Physics, Huazhong University of Science and Technology. 
\end{acknowledgments}


\begin{thebibliography}{99}
\bibitem{am54} Magnon lifetime is estimated to be on the order of 10 ps based on the Raman
spectra of two-dimensional collinear antiferromagnets.
\end{thebibliography}
\end{document}


\title{Supplemental Materials for ``Efficient Spin Seebeck and Spin Nernst Effects of Magnons in Altermagnets"}
       \affiliation{National Laboratory of Solid State Microstructures, School of Physics, Collaborative Innovation Center of Advanced Microstructures, Nanjing University, Nanjing 210093, China} 
        \affiliation{School of Physics, Huazhong University of Science and Technology, Wuhan 430074, China}
        \affiliation {Ningbo Institute of Materials Technology and Engineering, Chinese Academy of Sciences, Ningbo 315201, China} 
        \affiliation{Yongjiang Laboratory, Ningbo 315202, China}
        
        \author{Qirui Cui}  
        \affiliation{National Laboratory of Solid State Microstructures, School of Physics, Collaborative Innovation Center of Advanced Microstructures, Nanjing University, Nanjing 210093, China}
        \affiliation {Ningbo Institute of Materials Technology and Engineering, Chinese Academy of Sciences, Ningbo 315201, China} 
        \affiliation{Yongjiang Laboratory, Ningbo 315202, China}
        
        \author{Bowen Zeng}
        \affiliation{School of Physics, Huazhong University of Science and Technology, Wuhan 430074, China}
        
        \author{Tao Yu}       
        \email{taoyuphy@hust.edu.cn}
        \affiliation{School of Physics, Huazhong University of Science and Technology, Wuhan 430074, China}
        
        \author{Hongxin Yang}
        \email{hongxin.yang@nju.edu.cn} 
        \address{National Laboratory of Solid State Microstructures, School of Physics, Collaborative Innovation Center of Advanced Microstructures, Nanjing University, Nanjing 210093, China}
        
        \author{Ping Cui}
        \affiliation {Ningbo Institute of Materials Technology and Engineering, Chinese Academy of Sciences, Ningbo 315201, China} 
        \affiliation{Yongjiang Laboratory, Ningbo 315202, China}

\maketitle

\section{SPIN-SPLIT DISPERSION OF ALTERMAGNETIC MAGNONS}

The two-dimensional altermagnetic square lattice contains interpolated ``a" and ``b" sublattices, as illustrated in the main text. The spin Hamiltonian reads 
\begin{align} 
\hat{H}&=J_1 \sum_{\langle i, j\rangle} \hat{\bf {S}}_{a i}\cdot \hat{\bf S}_{b j}\nonumber\\
&+J_2 \sum_{\langle i_x, j_x\rangle} \hat{\bf{S}}_{a i}\cdot \hat{\bf{S}}_{a j}\nonumber
+J_2' \sum_{\langle i_y, j_y\rangle} \hat {\bf{S}}_{a i}\cdot \hat {\bf{S}}_{a j}+K \sum_i\left(\hat{S}_{a i}^z\right)^2 \nonumber\\
&+J_2' \sum_{\langle i_x, j_x\rangle} \hat {\bf{S}}_{b i}\cdot \hat {\bf{S}}_{b j}+J_2 \sum_{\langle i_y, j_y\rangle} \hat {\bf{S}}_{b i}\cdot \hat{\bf{S}}_{b j}+K \sum_i\left(\hat{S}_{b i}^z\right)^2,
\label{eS1} 
\end{align}
where the first term $J_1>0$ describes the antiferromagnetic spin interaction between``a" and ``b" sublattices, and the second and third terms $\{J_2, J_2'\}<0 $ contain the ferromagnetic spin interactions within the sublattices. $\langle i, j \rangle$, $\langle i_x, j_x \rangle$, and $\langle i_y ,j_y \rangle$ indicate the nearest-neighboring coupling between ``a" and ``b" sites along the $diagonal$, $\hat{\bf{x}}$, and $\hat{\bf{y}}$ direction, respectively. The anisotropy constant $K<0$ favors out-of-plane magnetization. 
In terms of the bosonic operators $\hat{a}_i$ and $\hat{b}_i$ for the sublattices ``a" and ``b", the Holstein-Primakoff transformation \cite{sm1} in the linear regime reads
\begin{subequations}
\begin{align}
\hat{S}_{a i}^{+} \approx \sqrt{2S} \hat{a}_i,~~~\hat{S}_{a i}^{-} \approx \sqrt{2S} \hat{a}_i^{\dagger},~~~ \hat{S}_{ai}^{z}=S-{\hat{a}_i}^{\dagger}{\hat{a}_i},\\
\hat{S}_{b j}^{+} \approx \sqrt{2S} \hat{b}_j^{\dagger},~~~\hat{S}_{b j}^{-} \approx \sqrt{2S} \hat{b}_j,~~~\hat{S}_{bj}^{z}={\hat{b}_j}^{\dagger}{\hat{b}_j}-S,
\end{align}
\end{subequations}
where
$\hat{S}_{i}^{\pm }=\hat{S}_{i}^{x}\pm i\hat{S}_{i}^{y}$. Disregarding the zero-point energy, the spin Hamiltonian (\ref{eS1}) becomes
\begin{align} \hat{H} & =J^*S\left(\sum_i \hat{a}_i^{\dagger} \hat{a}_i+\sum_j \hat {b}_j^{\dagger} \hat{b}_j\right)+2J_1S\sum_{\langle i, j \rangle}\left(\hat{a}_i \hat{b}_j+\hat{a}_i^{\dagger} \hat{b}_j^{\dagger}\right) \nonumber\\ & + J_2 S\sum_{\langle i_x, j_x \rangle}\left(\hat{a}_{i} \hat{a}_{j}^{\dagger}+\hat{a}_{i}^{\dagger} \hat{a}_{j}\right)+ J_2^{\prime}S\sum_{\langle i_y, j_y \rangle}\left(\hat{a}_{i} \hat{a}_{j}^{\dagger}+\hat{a}_{i}^{\dagger} \hat{a}_{j}\right) \nonumber\\ & + J_2^{\prime}S\sum_{\langle i_x, j_x \rangle}\left(\hat{b}_{i} \hat{b}_{j}^{\dagger}+\hat{b}_{i}^{\dagger} \hat{b}_{j}\right)+ J_2 S\sum_{\langle i_y, j_y \rangle}\left(\hat{b}_{i} \hat{b}_{j}^{\dagger}+\hat{b}_{i}^{\dagger} \hat{b}_{j}\right),
\end{align}
where the on-site energy $J^*=4J_1-2J_2-2J_2^{\prime}-2K$. With $\hat{a}_i=\sum_{\bf k} e^{i {\bf k} \cdot {\bf R}_i} \hat{a}_{\bf k}$ and $\hat{b}_j=\sum_{\bf k} e^{-i {\bf k} \cdot {\bf R}_j} \hat{b}_{\bf k}$, we transform to the wave-vector space: 
\begin{align}
\hat{H}=\sum_{\bf k}\left(\hat{a}_{\bf k}^{\dagger}, \hat{b}_{\bf k}\right) \hat{H}_{\bf k} \left(\begin{array}{l}\hat{a}_{\bf k} \\ \hat{b}_{\bf k}^{\dagger}\end{array}\right),~~~ \hat{H}_{\bf k} = S \left(\begin{array}{cc}J^*+2 J_2 \gamma_1+2 J_2^{\prime} \gamma_2 & 4 J_1 \gamma_3 \\ 4 J_1 \gamma_3 & J^*+2 J_2^{\prime} \gamma_1+2 J_2 \gamma_2\end{array}\right) ,\label{eS3}
\end{align}
where $\gamma_1=\cos(k_xa)$, $\gamma_2=\cos(k_ya)$, and $\gamma_3=\cos(k_xa/2)\cos(k_ya/2)$.  

We find the magnon modes of Eq.~(\ref{eS3}) by the Bogoliubov transformation 
\begin{align}
\left(\begin{array}{cc}\hat{a}_{\bf k} \\ \hat{b}_{\bf k}^{\dagger} \end{array}\right)=T_{\bf k} \left(\begin{array}{cc}\hat{\alpha}_{\bf k} \\ \hat{\beta}_{\bf k}^{\dagger} \end{array}\right), \end{align}
where the paraunitary matrix
\begin{align}
T_{\bf k}=\left(\begin{array}{cc}u_{\bf k} & v_{\bf k} \\ v_{\bf k} & u_{\bf k}\end{array}\right),
\end{align}
leading to $\left[J^*+\left(J_2+ J_2^{\prime}\right)\left(\gamma_1+\gamma_2\right)\right]u_{\bf k}v_{\bf k}+2J_1\gamma_3(u_{\bf k}^2+v_{\bf k}^2)= 0$. On the other hand, the bosonic commutation relation for the magnon operators leads to the relation $T_{\bf k}^{\dagger} \hat{\sigma}_z T_{\bf k} = \hat{\sigma}_z$ with $\hat{\sigma}_z$ = diag$(1,-1)$ that requests $u_{\bf k}^2-v_{\bf k}^2=1$. We thus obtain
\begin{align}
 u_{\bf k}=\sqrt{(\Delta^{-1}+1)/2},~~~ 
 v_{\bf k}=-\sqrt{(\Delta^{-1}-1)/2}, ~~~
\Delta=\sqrt{1-{\gamma_e}^2}, 
\end{align}
where the effective form factor $\gamma_e= {4J_1\gamma_3}/[{4J_1-2K-(J_2+J_2^{\prime})(2-\gamma_1-\gamma_2)}]$.
The magnon Hamiltonian contains the $\alpha$- and $\beta$-modes with 
\begin{align} 
\hat{H}= \sum_{\bf k}\left(\hat{\alpha}_{\bf k}^{\dagger}, \hat{\beta}_{\bf k}\right) \left(\begin{array}{cc} \hbar\omega_{\bf k}^\alpha & 0 \\ 0 & \hbar\omega_{\bf k}^\beta \end{array}\right) \left(\begin{array}{l}\hat{\alpha}_{\bf k} \\ \hat{\beta}_{\bf k}^{\dagger}\end{array}\right),
\end{align}
where the eigenfrequency of $\alpha$- and $\beta$-modes
 \begin{align}
&\omega_{\bf k}^\alpha=\frac{S}{\hbar} \left[(J_2-J_2^{\prime})\left(\gamma_1-\gamma_2\right)+\frac{4J_1\gamma_3 \Delta}{\gamma_e}\right],\nonumber\\
&\omega_{\bf k}^\beta=\frac{S}{\hbar} \left[(J_2^{\prime}-J_2)\left(\gamma_1-\gamma_2\right)+\frac{4J_1\gamma_3 \Delta}{\gamma_e}\right],
\end{align}
are not degenerate since $J_2\ne J_2'$.

The $\alpha$ and $\beta$ modes turn out to carry opposite ``spins". Via the Holstein-Primakoff and Bogoliubov transformations, the $z$-component of the total spin operator reads
\begin{align}
\nonumber
\hat{S}^z &\equiv \sum_i\left(\hat{S}_{ai}^z+\hat{S}_{bi }^z\right)=\sum_i\left(-\hat{a}_{i}^{\dagger}\hat{a}_{i}+\hat{b}_{i }^{\dagger}\hat{b}_{i}\right)
=\sum_{\bf k}\left(-\hat{a}_{\bf k}^{\dagger}\hat{a}_{\bf k}+\hat{b}_{\bf k}^{\dagger}\hat{b}_{\bf k}\right)\\
&=\sum_{\bf k}\left(-\hat{\alpha}_{\bf k}^{\dagger}\hat{\alpha}_{\bf k}+\hat{\beta}_{\bf k}^{\dagger}\hat{\beta}_{\bf k}\right).   
\end{align}
Consequently, the spin angular momentum (with the unit of $\hbar$) along the $\hat{\bf{z}}$-direction for any $\alpha$ and $\beta$ modes
\begin{subequations}
\begin{align}
&\langle0|\hat{\alpha}_{\bf k} \hat{S}^z \hat{\alpha}_{\bf k}^{\dagger}|0\rangle=-1,\\
&\langle 0|\hat{\beta}_{\bf k} \hat{S}^z \hat{\beta}_{\bf k}^{\dagger}|0\rangle=1,
\end{align}
\end{subequations}
where $|0\rangle$ indicates the magnon vacuum. 

\section{Details of First-principles calculations}
\begin{figure}[h]
\includegraphics[width=0.9\linewidth]{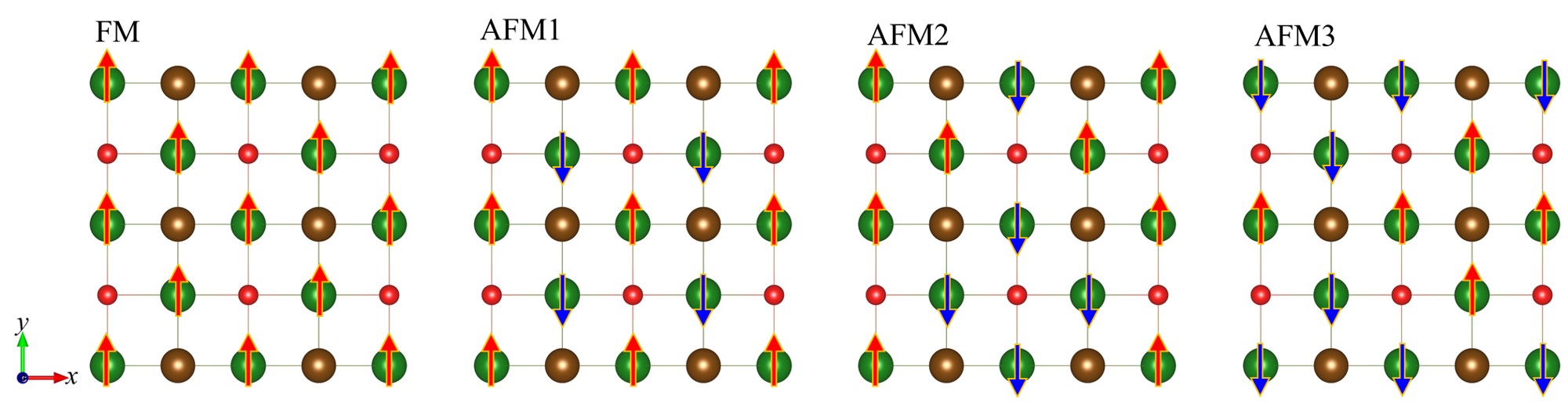}
\caption{Spin configurations, indicated by red and blue arrows, for FM, AFM1, AFM2, and AFM3 that are chosen for extracting exchange couplings between Cr atoms in $\rm Cr_2Te_2O$ and $\rm Cr_2Se_2O$ monolayers.}
\label{FigS1}
\end{figure}

The first-principles calculations are implemented in the Vienna $ab$ $initio$ Simulation Package (VASP), based on the density functional theory (DFT)~\cite{sm2, sm3, sm4}. The exchange correlation functional is treated within the generalized gradient approximation (GGA) of Perdew-Burke-Ernzerhof form~\cite{sm5}. For describing the strong correlation effects of 3$d$ electrons, the GGA+U method is applied with an effective Hubbard $U$ parameter that is chosen to be 4 eV for transition mental elements. $U_{\rm eff}$=3, 5 eV are also tested, and the main findings about the material candidate are robust against $U_{\rm eff}$ values. The crystal structure is fully relaxed until the Hellmann-Feynman force acting on each atom is less than $10^{-3}$ eV/Å. The plane-wave cutoff energy is set to be 420 eV and the two-dimensional Brillouin zone is sampled using a $\Gamma$-centered 20×20×1 Monkhorst-pack $k$ mesh. In order to avoid interactions between adjacent layers, the vacuum space is set to be larger than 15 \AA. 
The finite displacement method is applied for obtaining the phonon band dispersion in a 4×4×1 supercell, based on PHONOPY code \cite{sm6}. $Ab~initio$ molecular dynamics simulations are performed in the canonical NVT particles ensemble used with Nos\'e thermostat \cite{sm7}. Such simulations are conducted for 10 ps at 500 K. The energy mapping methods are applied for extracting magnetic parameters \cite{sm8, sm9}. Specifically, the self-consistent energies of a 2×2×1 supercell with spin configurations FM, AFM1, AFM2, and AFM3 (Fig. \ref{FigS1}) are calculated, and accordingly, the nearest-neighboring exchange coupling $J_1$ and next-nearest-neighboring exchange coupling $J_2(J_2^{\prime})$ reads $J_1=\left( E_{FM}-E_{AFM1} \right)/32S^{2}$, $J_2=\left(E_{AFM1}+E_{FM}-2E_{AFM2} \right)/32S^{2}$ and $J_2^{\prime}=\left(E_{AFM1}+E_{FM}-2E_{AFM3}\right)/32S^{2}$. The magnetic anisotropy is estimated by the energy difference of the unit cell with out-of-plane magnetization along $\hat{\bf z}$ and in-plane magnetization along $\hat{\bf x}$. $\rm Cr_2Se_2O$ possesses the same structure with $\rm Cr_2Te_2O$, and the optimizted lattice constant $a = 4.04$ \AA. The bonding angle of Cr–Se–Cr and Cr-O-Cr equal $\rm 100^{\circ}$ and $\rm 180^{\circ}$, respectively.

\section{Kubo theory for spin and thermal conductivities}

The magnon thermal conductivity tensor is defined via the relation
\begin{align}
\left\langle \hat{j}_m^{z}\right\rangle=\sigma_{mn} (-\partial_n T), 
\label{es12}
\end{align}
where $\left\langle \hat{j}_m^{z}\right\rangle$ represents the spin current density flowing along the $\hat{\bf m}$-direction and $\partial_n T$ represents the temperature gradient along $n$-direction. In terms of Kubo formula~\cite{sm10,sm11},
\begin{align}
 \left\langle \hat{j}_m^z\right\rangle=\lim _{\Omega \rightarrow 0}\left\{\left[\Pi_{m n}^R(\Omega)-\Pi_{m n}^R(0)\right] / \mathrm{i} \Omega\right\} \frac{-\hbar \partial_n T}{AT}, \label{es13}
 \end{align} 
 where $\Pi_{m n}^R(\Omega)$ is the retarded current-current correlation function, $\Omega$ is the frequency, and $A$ represents the sample area. Substituting
(\ref{es12}) into (\ref{es13}) yields 
 \begin{align}
 \sigma_{mn}=\frac{\hbar}{AT}\lim _{\Omega \rightarrow 0}\left\{\left[\Pi_{mn}^R(\Omega)-\Pi_{mn}^R(0)\right] / \mathrm{i} \Omega\right\}. \label{es14}
 \end{align}

$\Pi_{mn}^R(\Omega)$ is further evaluated in the Matsubara representation \cite{sm11}:
\begin{align}
\nonumber
\Pi_{mn}^R(\Omega)&=\Pi_{mn}(\mathrm{i} \Omega \rightarrow \Omega+\mathrm{i} 0^+),\\
\Pi_{mn}(\mathrm{i} \Omega)&=-\int_0^{\beta} \mathrm{e}^{\mathrm{i} \tau \Omega}\left\langle \hat{\mathcal{T}} \hat{J}_m (\tau) \hat{Q}_n(0)\right\rangle \mathrm{d}\tau, \label{es16}
\end{align}
where $\beta=1/(k_{B}T)$, $\tau$ is the imaginary time, $\hat{\mathcal{T}}$ is the time-ordering operator, and $\Omega=2\pi l/\beta$ is the bosonic Matsubara frequency with integer $l$. In terms of magnon basis, the total spin and thermal current reads \cite{sm12, sm13, sm14} 
\begin{subequations}
\begin{align}
&\hat{J}_m=\sum_{\bf{k}} \hat{\Psi}_{\bf{k}}^{\dagger} \hat{\mathcal{J}}_{m, \bf{k}} \hat{\Psi}_{\bf{k}},~~~\hat{\mathcal{J}}_{m, \bf{k}}=\frac{1}{2}  T_{\bf {k}}^{\dagger} \left( \hat{v}_{m, \bf {k}} \hat{\sigma}_z \hat{S}^z + \hat{S}^z \hat{\sigma}_z \hat{v}_{m, \bf {k}} \right) T_{\bf {k}} = -T_{\bf {k}}^{\dagger} \left( \partial_{k_m}\hat{H}_{\bf k} \right) T_{\bf {k}},\\
&\hat{Q}_n=\sum_{\bf{k}} \hat{\Psi}_{\bf{k}}^{\dagger} \hat{\mathcal{Q}}_{n, \bf{k}} \hat{\Psi}_{\bf{k}},~~~\hat{\mathcal{Q}}_{n, \bf{k}}=\frac{1}{2}  T_{\bf {k}}^{\dagger} \left( \hat{v}_{n, \bf {k}} \hat{\sigma}_z \hat{H}_{\bf k} + \hat{H}_{\bf k} \hat{\sigma}_z \hat{v}_{n, \bf {k}} \right) T_{\bf {k}},
\end{align}
\end{subequations}
where velocity operator $\hat{v}_{m, \bf {k}}= (\partial_{k_m}\hat{H}_{\bf k}) / \hbar$. Following Ref.~\cite{sm13}, we transform (\ref{es16}) to wave-vector space:
\begin{align}
\Pi_{mn}(\mathrm{i} \Omega)&=-\int_0^{\beta} \mathrm{e}^{\mathrm{i} \tau \Omega} \sum_{\bf k, k'} \sum_{a,b,c,d} ( \hat{\mathcal{J}}_{m, \bf{k}})_{ab} ( \hat{\mathcal{Q}}_{n, \bf{k'}})_{cd} \left\langle \hat{\mathcal{T}} \hat{\Psi}_{a, \bf{k}}^{\dagger}(\tau) \hat{\Psi}_{b, \bf{k}}(\tau) \hat{\Psi}_{c, \bf{k'}}^{\dagger}(0) \hat{\Psi}_{d, \bf{k'}}(0)  \right\rangle \mathrm{d}\tau \nonumber \\
&=-\int_0^{\beta} \mathrm{e}^{\mathrm{i} \tau \Omega} \sum_{\bf k, k'} \sum_{a,b,c,d} ( \hat{\mathcal{J}}_{m, \bf{k}})_{ab} ( \hat{\mathcal{Q}}_{n, \bf{k'}})_{cd} \nonumber \\ & \left( \left\langle \hat{\mathcal{T}} \hat{\Psi}_{a, \bf{k}}^{\dagger}(\tau) \hat{\Psi}_{c, \bf{k'}}^{\dagger}(0) \right\rangle \left\langle \hat{\mathcal{T}} \hat{\Psi}_{b, \bf{k}}(\tau) \hat{\Psi}_{d, \bf{k'}}(0) \right\rangle + \left\langle \hat{\mathcal{T}} \hat{\Psi}_{a, \bf{k}}^{\dagger}(\tau) \hat{\Psi}_{d, \bf{k'}}(0) \right\rangle \left\langle \hat{\mathcal{T}} \hat{\Psi}_{b, \bf{k}}(\tau) \hat{\Psi}_{c, \bf{k'}}^{\dagger}(0) \right\rangle \right) \mathrm{d}\tau,
\end{align}
where $a, b, c, d \in \{\alpha,\beta\}$ indicates the band index.
For the bosonic operators with diagonal basis, the time-ordering ($\tau>0$) average of equilibrium states follows
\begin{subequations}
\begin{align}
&\left\langle \hat{\mathcal{T}} \hat{\alpha}_{\bf{k}}^{\dagger}(\tau'+\tau) \hat{\beta}_{\bf{k'}}(\tau') \right\rangle = \left\langle \hat{\mathcal{T}} \hat{\alpha}_{\bf{k}}(\tau'+\tau) \hat{\beta}_{\bf{k'}}^{\dagger}(\tau') \right\rangle = 0, \\
&\left\langle \hat{\mathcal{T}} \hat{\alpha}_{\bf{k}}^{\dagger}(\tau'+\tau) \hat{\alpha}_{\bf{k'}}(\tau') \right\rangle = \delta_{\bf k,k'} g\left(\hbar \omega_{\bf k}^\alpha \right) e^{\tau \hbar \omega_{\bf k}^\alpha},\nonumber\\
&\left\langle \hat{\mathcal{T}} \hat{\alpha}_{\bf{k}}(\tau'+\tau) \hat{\alpha}_{\bf{k'}}^{\dagger}(\tau') \right\rangle = -\delta_{\bf k,k'} g\left(-\hbar \omega_{\bf k}^\alpha \right) e^{-\tau \hbar \omega_{\bf k}^\alpha},
\end{align}
\end{subequations}
with Bose-Einstein distribution $g\left(\hbar \omega_{\bf k}^\alpha \right)= (e^{\hbar \omega_{\bf k}^\alpha/k_{B}T}-1)^{-1}$.
Accordingly, 
\begin{align}
\Pi_{mn}(\mathrm{i} \Omega) & =-\int_0^{\beta} \mathrm{e}^{\mathrm{i} \tau \Omega} \sum_{\bf k, k'} \sum_{a,b,c,d} \left( \hat{\mathcal{J}}_{m, \bf{k}} \right)_{ab} \left( \hat{\mathcal{Q}}_{n, \bf{k'}} \right)_{cd} \left\langle \hat{\mathcal{T}} \hat{\Psi}_{a, \bf{k}}^{\dagger}(\tau) \hat{\Psi}_{d, \bf{k'}}(0) \right\rangle \left\langle \hat{\mathcal{T}} \hat{\Psi}_{b, \bf{k}}(\tau) \hat{\Psi}_{c, \bf{k'}}^{\dagger}(0) \right\rangle \mathrm{d}\tau \nonumber \\&= \sum_{\bf {k}} \sum_{a, b} \left( \hat{\mathcal{J}}_{m, \bf{k}} \right)_{a b} \left( \hat{\mathcal{Q}}_{n, \bf{k}} \right)_{b a} \left( \hat{\sigma}_z \right)_{a a} \left( \hat{\sigma}_z \right)_{b b} \frac{g\left[\left(\hat{\sigma}_z \mathcal{E}_{\bf {k}}\right)_{b b}\right]-g\left[\left(\hat{\sigma}_z \mathcal{E}_{\boldsymbol{k}}\right)_{a a}\right]}{\left(\hat{\sigma}_z \mathcal{E}_{\bf {k}}\right)_{a a}-\left(\hat{\sigma}_z \mathcal{E}_{\bf {k}}\right)_{b b}+\mathrm{i} \Omega}, \label{es20}
\end{align}
where eigenmatrix
\begin{align}
\mathcal{E}_{\bf {k}} = \left(\begin{array}{cc} \hbar\omega_{\bf k}^\alpha & 0 \\ 0 & \hbar\omega_{\bf k}^\beta \end{array}\right).
\end{align}
 We substitute (\ref{es20}) into (\ref{es14}) 
 with $i0^{+} \rightarrow i\Gamma$, where $\Gamma$ governs quasiparticle broadening, and obtain  the magnon thermal conductivity
 \begin{align}
 \sigma_{mn}=-\frac{\hbar}{A T} \sum_{\bf k} \sum_{a, b}  \operatorname{Re}\left[\left( \hat{\mathcal{J}}_{m, \bf{k}} \right)_{ab} \left( \hat{\mathcal{Q}}_{n, \bf{k}} \right)_{ba}\right] \frac{g\left[\left(\hat{\sigma}_z \mathcal{E}_{\bf {k}}\right)_{a a}\right]-g\left[\left(\hat{\sigma}_z \mathcal{E}_{\bf {k}}\right)_{b b}\right]}{\left(\hat{\sigma}_z \mathcal{E}_{\bf {k}}\right)_{a a}-\left(\hat{\sigma}_z \mathcal{E}_{\bf {k}}\right)_{b b}}
 \frac{\Gamma \left(\hat{\sigma}_z\right)_{a a} \left(\hat{\sigma}_z\right)_{b b}}{\Gamma^2+\left[\left(\hat{\sigma}_z \mathcal{E}_{\bf {k}}\right)_{a a}-\left(\hat{\sigma}_z \mathcal{E}_{\bf {k}}\right)_{b b}\right]^2}.    
 \end{align}
 By defining magnon lifetime $\tau_0={\hbar}/{\Gamma}$ and disregarding magnon-magnon interaction at low temperature~\cite{sm14, sm15}, we arrive at 
\begin{align}
 \sigma_{mn} = &- \frac{\tau_0}{A k_{B} T^2} \sum_{\bf k} \left( T_{\bf {k}}^{\dagger} (\partial_{k_m}\hat{H}_{\bf k})  T_{\bf {k}} \right)_{\alpha \alpha}  \left( T_{\bf {k}}^{\dagger} (\partial_{k_n}\hat{H}_{\bf k})  T_{\bf {k}} \right)_{\alpha \alpha} \omega_{\bf k}^{\alpha} \frac{e^{\hbar \omega_{\bf k}^{\alpha}  /k_{B}T}}{\left(e^{\hbar \omega_{\bf k}^{\alpha} /k_{B}T}-1\right)^2} \nonumber \\
 & + \frac{\tau_0}{A k_{B} T^2} \sum_{\bf k} \left( T_{\bf {k}}^{\dagger} (\partial_{k_m}\hat{H}_{\bf k})  T_{\bf {k}} \right)_{\beta \beta}  \left( T_{\bf {k}}^{\dagger} (\partial_{k_n}\hat{H}_{\bf k})  T_{\bf {k}} \right)_{\beta \beta} \omega_{\bf k}^{\beta} \frac{e^{-\hbar \omega_{\bf k}^{\beta}  /k_{B}T}}{\left(e^{-\hbar \omega_{\bf k}^{\beta} /k_{B}T}-1\right)^2}.
 \label{es23}
\end{align}

A similar procedure is applied for the calculation of thermal conductivity $\kappa_{mn}$. Different from spin conductivity,  
here $\Pi_{mn}(\mathrm{i}\Omega)$ is rewritten as
\begin{align}
\Pi_{mn}(\mathrm{i} \Omega)=-\int_0^{\beta} \mathrm{e}^{\mathrm{i} \tau \Omega}\left\langle \hat{\mathcal{T}} \hat{Q}_m (\tau) \hat{Q}_n(0)\right\rangle \mathrm{d}\tau. \label{es26}
\end{align} 
Transforming (\ref{es26}) to wave-vector space, 
\begin{align}
\Pi_{mn}(\mathrm{i} \Omega) = \sum_{\bf {k}} \sum_{a, b} \left( \hat{\mathcal{Q}}_{m, \bf{k}} \right)_{a b} \left( \hat{\mathcal{Q}}_{n, \bf{k}} \right)_{b a} \left( \hat{\sigma}_z \right)_{a a} \left( \hat{\sigma}_z \right)_{b b} \frac{g\left[\left(\hat{\sigma}_z \mathcal{E}_{\bf {k}}\right)_{b b}\right]-g\left[\left(\hat{\sigma}_z \mathcal{E}_{\boldsymbol{k}}\right)_{a a}\right]}{\left(\hat{\sigma}_z \mathcal{E}_{\bf {k}}\right)_{a a}-\left(\hat{\sigma}_z \mathcal{E}_{\bf {k}}\right)_{b b}+\mathrm{i} \Omega}. \label{es28}
\end{align}
Substituting (\ref{es28}) into (\ref{es14}), we obtain 
\begin{align}
 \kappa_{mn}= & \frac{\tau_0}{A k_{B} T^2} \sum_{\bf k} \left( T_{\bf {k}}^{\dagger} (\partial_{k_m}\hat{H}_{\bf k})  T_{\bf {k}} \right)_{\alpha \alpha}  \left( T_{\bf {k}}^{\dagger} (\partial_{k_n}\hat{H}_{\bf k})  T_{\bf {k}} \right)_{\alpha \alpha} \left( \omega_{\bf k}^{\alpha} \right)^2 \frac{e^{\hbar  \omega_{\bf k}^{\alpha}  /k_{B}T}}{\left(e^{\hbar \omega_{\bf k}^{\alpha} /k_{B}T}-1\right)^2}  \nonumber \\
 + & \frac{\tau_0}{A k_{B} T^2} \sum_{\bf k} \left( T_{\bf {k}}^{\dagger} (\partial_{k_m}\hat{H}_{\bf k})  T_{\bf {k}} \right)_{\beta \beta}  \left( T_{\bf {k}}^{\dagger} (\partial_{k_n}\hat{H}_{\bf k})  T_{\bf {k}} \right)_{\beta \beta} \left( \omega_{\bf k}^{\beta} \right)^2 \frac{e^{-\hbar \omega_{\bf k}^{\beta}  /k_{B}T}}{\left(e^{-\hbar \omega_{\bf k}^{\beta} /k_{B}T}-1\right)^2}. 
\end{align}

\section{Boltzmann transport theory for spin and thermal conductivities}

With the temperature gradient along the $\hat{\bf n}$ direction $\partial_{n} T$, the Boltzmann
equation in the relaxation time $\tau_0$ approximation is given by 
\begin{align}
\frac{g_{neq}-g_{eq}}{\tau_0}=-v_{n} \partial_{n} T \frac{\partial g_{eq}}{\partial T}=-v_{n} \partial_{n} T \frac{E}{k_{B}T^2} \frac{e^{E /k_{B}T}}{\left(e^{E /k_{B}T}-1\right)^2},
\end{align}
where the equilibrium thermal distribution $g_{eq}= (e^{E/k_{B}T}-1)^{-1}$. 
The consequent spin-up current density  along the $\hat{\bf m}$ direction reads 
\begin{align}
\left\langle \hat{j}_m^{z}\right\rangle= \frac{v_{m} \hbar}{A}(g_{neq}-g_{eq})= -\frac{\tau_0 E}{A k_{B}T^2} \hbar v_{m} v_{n} \partial_{n} T \frac{e^{E /k_{B}T}}{\left(e^{E /k_{B}T}-1\right)^2}.
\end{align}
Accordingly, 
\begin{subequations}
\begin{align}
& \sigma_{mn} = \sigma_{mn}^{\alpha} + \sigma_{mn}^{\beta}, \\
& \sigma_{mn}^{\alpha} = - \frac{\tau_0}{A k_{B} T^2} \sum_{\bf k} \left( T_{\bf {k}}^{\dagger} (\partial_{k_m}\hat{H}_{\bf k})  T_{\bf {k}} \right)_{\alpha \alpha}  \left( T_{\bf {k}}^{\dagger} (\partial_{k_n}\hat{H}_{\bf k})  T_{\bf {k}} \right)_{\alpha \alpha} \omega_{\bf k}^{\alpha} \frac{e^{\hbar \omega_{\bf k}^{\alpha}  /k_{B}T}}{\left(e^{\hbar \omega_{\bf k}^{\alpha} /k_{B}T}-1\right)^2}, \\
& \sigma_{mn}^{\beta} = \frac{\tau_0 }{A k_{B} T^2} \sum_{\bf k} \left( T_{\bf {k}}^{\dagger} (\partial_{k_m}\hat{H}_{\bf k})  T_{\bf {k}} \right)_{\beta \beta}  \left( T_{\bf {k}}^{\dagger} (\partial_{k_n}\hat{H}_{\bf k})  T_{\bf {k}} \right)_{\beta \beta} \omega_{\bf k}^{\beta} \frac{e^{\hbar \omega_{\bf k}^{\beta}  /k_{B}T}}{\left(e^{\hbar \omega_{\bf k}^{\beta} /k_{B}T}-1\right)^2}.
\end{align}
\end{subequations}
With the identity
\begin{align}
\frac{e^{E /k_{B}T}}{\left(e^{E /k_{B}T}-1\right)^2} = \frac{e^{-E /k_{B}T}}{\left(e^{-E /k_{B}T}-1\right)^2}, 
\end{align}
we find the same spin conductivity derived based on the Kubo formula by constant
lifetime approximation and restricting to intraband contribution [see Eq.~\ref{es23}]. 

The same approach is adopted for resolving the thermal conductivity by defining the thermal current density
\begin{align}
\left\langle \hat{j}_m^{t}\right\rangle= \frac{v_{m} E}{A}(g_{neq}-g_{eq})= -\frac{\tau_0 E^{2}}{A k_{B}T^2} v_{m} v_{n} \partial_{n} T \frac{e^{E /k_{B}T}}{\left(e^{E /k_{B}T}-1\right)^2}.
\end{align} 
Accordingly, 
\begin{align}
\kappa_{mn} & = \kappa_{mn}^{\alpha}+\kappa_{mn}^{\beta}, \nonumber \\
\kappa_{mn}^{\alpha} & = \frac{\tau_0}{A k_{B} T^2} \sum_{\bf k} \left( T_{\bf {k}}^{\dagger} (\partial_{k_m}\hat{H}_{\bf k})  T_{\bf {k}} \right)_{\alpha \alpha}  \left( T_{\bf {k}}^{\dagger} (\partial_{k_n}\hat{H}_{\bf k})  T_{\bf {k}} \right)_{\alpha \alpha} \left( \omega_{\bf k}^{\alpha} \right)^2 \frac{e^{\hbar  \omega_{\bf k}^{\alpha}  /k_{B}T}}{\left(e^{\hbar \omega_{\bf k}^{\alpha} /k_{B}T}-1\right)^2},  \nonumber \\
\kappa_{mn}^{\beta} & =\frac{\tau_0}{A k_{B} T^2} \sum_{\bf k} \left( T_{\bf {k}}^{\dagger} (\partial_{k_m}\hat{H}_{\bf k})  T_{\bf {k}} \right)_{\beta \beta}  \left( T_{\bf {k}}^{\dagger} (\partial_{k_n}\hat{H}_{\bf k})  T_{\bf {k}} \right)_{\beta \beta} \left( \omega_{\bf k}^{\beta} \right)^2 \frac{e^{\hbar \omega_{\bf k}^{\beta}  /k_{B}T}}{\left(e^{\hbar \omega_{\bf k}^{\beta} /k_{B}T}-1\right)^2}. 
\end{align}

\section{Properties of altermagnets and altermagnetic magnons}

\begin{figure}[htb]
\includegraphics[width=1\linewidth]{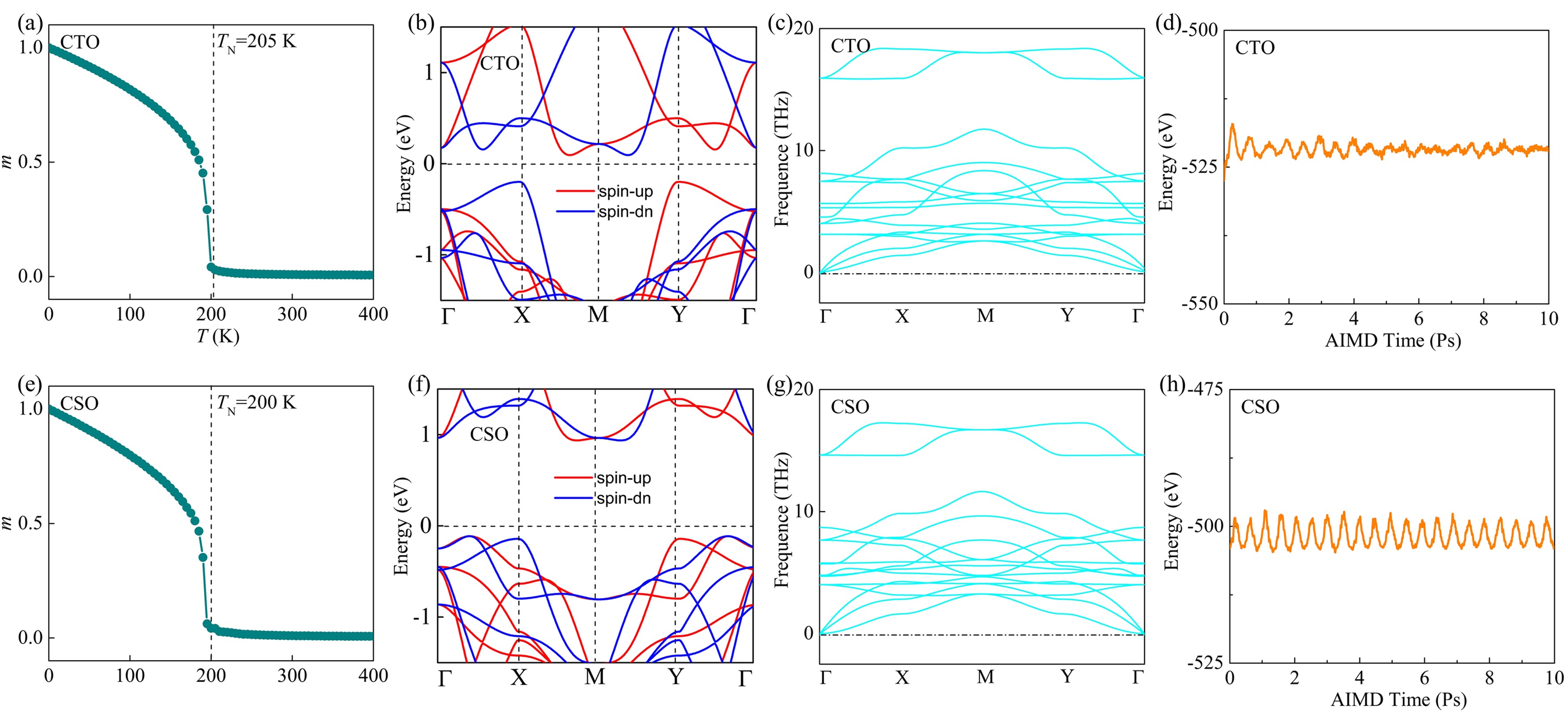}
\caption{(a) and (c) Monte Carlo simulations of temperature-dependent 
 N\'{e}el magnetization. Net magnetization of the antiferromagnetic system equals zero indicating the atomic spin of each sublattice is aligned. Accordingly, the normalized magnetization $m_1$ and $m_2$ of two sublattices are defined and $m$=($m_1$+$m_2$)/2 is applied to estimate the N\'eel temperature. (b) and (f) Spin-polarized band structure. The indirect band gap of $\rm Cr_2Te_2O$ and $\rm Cr_2Se_2O$ reach 0.30 and 1.05~eV, respectively. (c) and (g) Phonon band dispersions. (d) and (h) Evolution of total energy with temperature $T=500$~K during $ab~initio$ molecular dynamics simulations.}
\label{FigS2}
\end{figure}

\begin{figure}[h]
\includegraphics[width=1\linewidth]{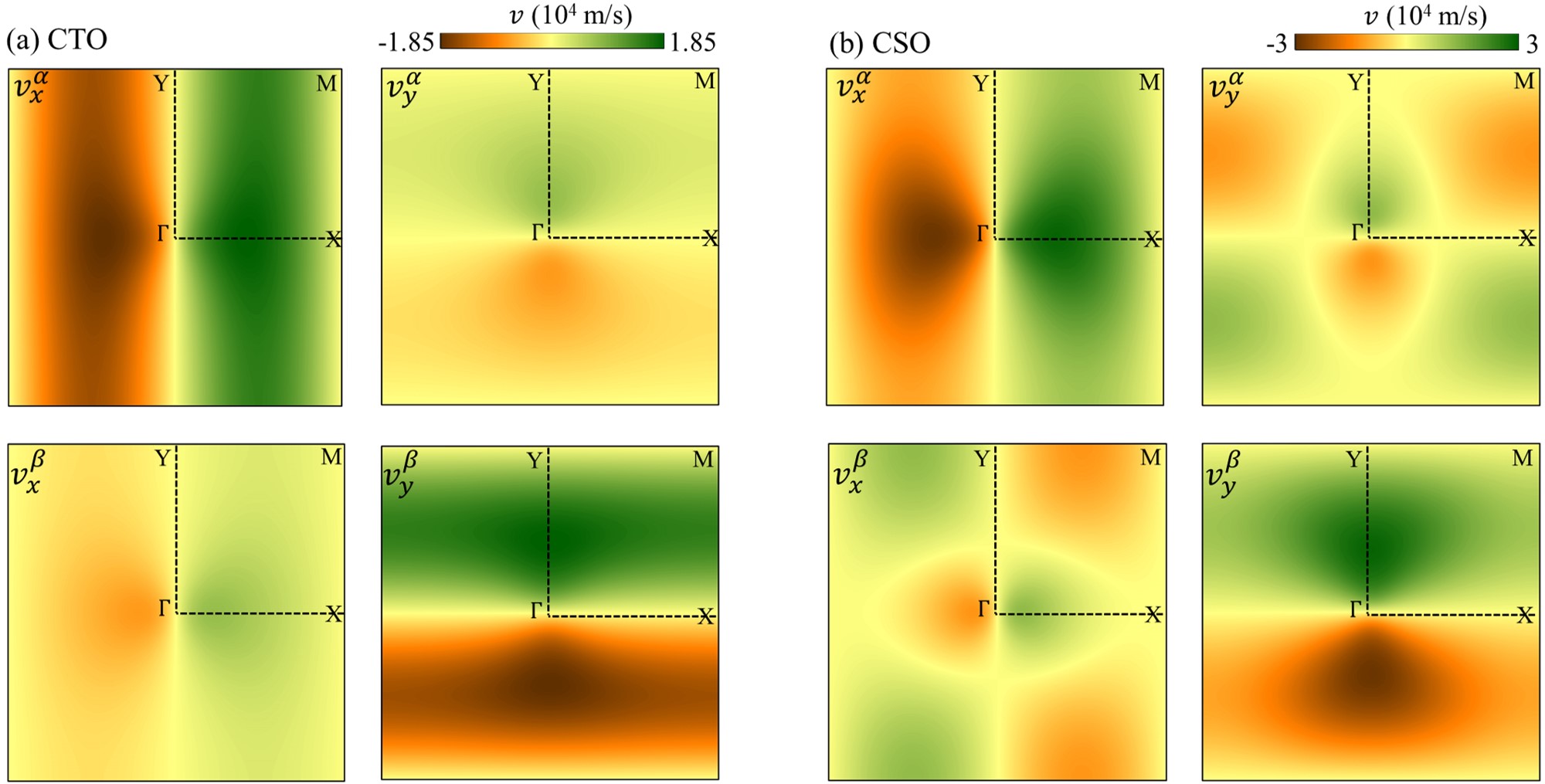}
\caption{The group velocity $v$ of magnons of (a) $\rm Cr_2Te_2O$ and (b) $\rm Cr_2Se_2O$. Specifically, $v_n^{\alpha}=(1/\hbar) (\!T_{\bf {k}}^{\dagger} (\partial_{k_n}\hat{H}_{\bf k})  T_{\bf {k}})_{\alpha \alpha}$ and $v_n^{\beta} = (1/\hbar) (\!T_{\bf {k}}^{\dagger} (\partial_{k_n}\hat{H}_{\bf k})  T_{\bf {k}})_{\beta \beta}$. Notably, the magnitude of $v$ reaches $10^{4}$ m/s. $v$ exhibits significant anisotropy in $\hat{\bf x}$ and $\hat{\bf y}$ directions.}
\label{FigS3}
\end{figure}

\begin{figure}[h]
\includegraphics[width=0.9\linewidth]{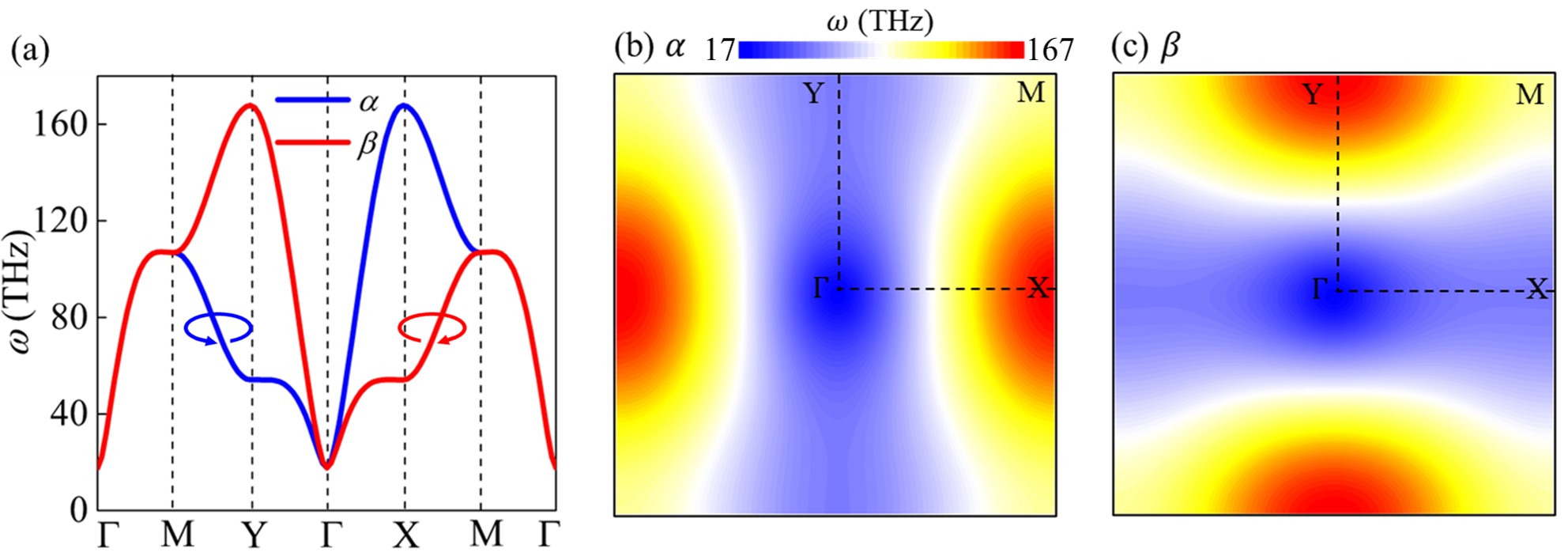}
\caption{(a) Spin-splitting magnon band structure and (b, c) magnon energy spectrum of $\rm Cr_2Se_2O$. Similar to $\rm Cr_2Te_2O$, anisotropic spin-momentum locking emerges, leading to the spin Seebeck and Nernst effects without the assistance of magnetic field and spin-orbit coupling. For $T=100$~K and $\tau_{0}=10$~ps, the maximum magnitude of $\sigma_{\parallel(\perp)}$ reaches  1.62~meV/K.}  
\label{FigS4}
\end{figure}

Figure~\ref{FigS2} shows the magnetic and structural properties of $\rm Cr_2Te_2O$ and $\rm Cr_2Se_2O$. The group velocity of altermagnetic magnons is shown in Fig.~\ref{FigS3}. We plot the magnon band dispersion of $\rm Cr_2Se_2O$ in Fig.~\ref{FigS4}.

\newpage
\section{Other potential candidates}

\begin{figure}[h]
\includegraphics[width=0.9\linewidth]{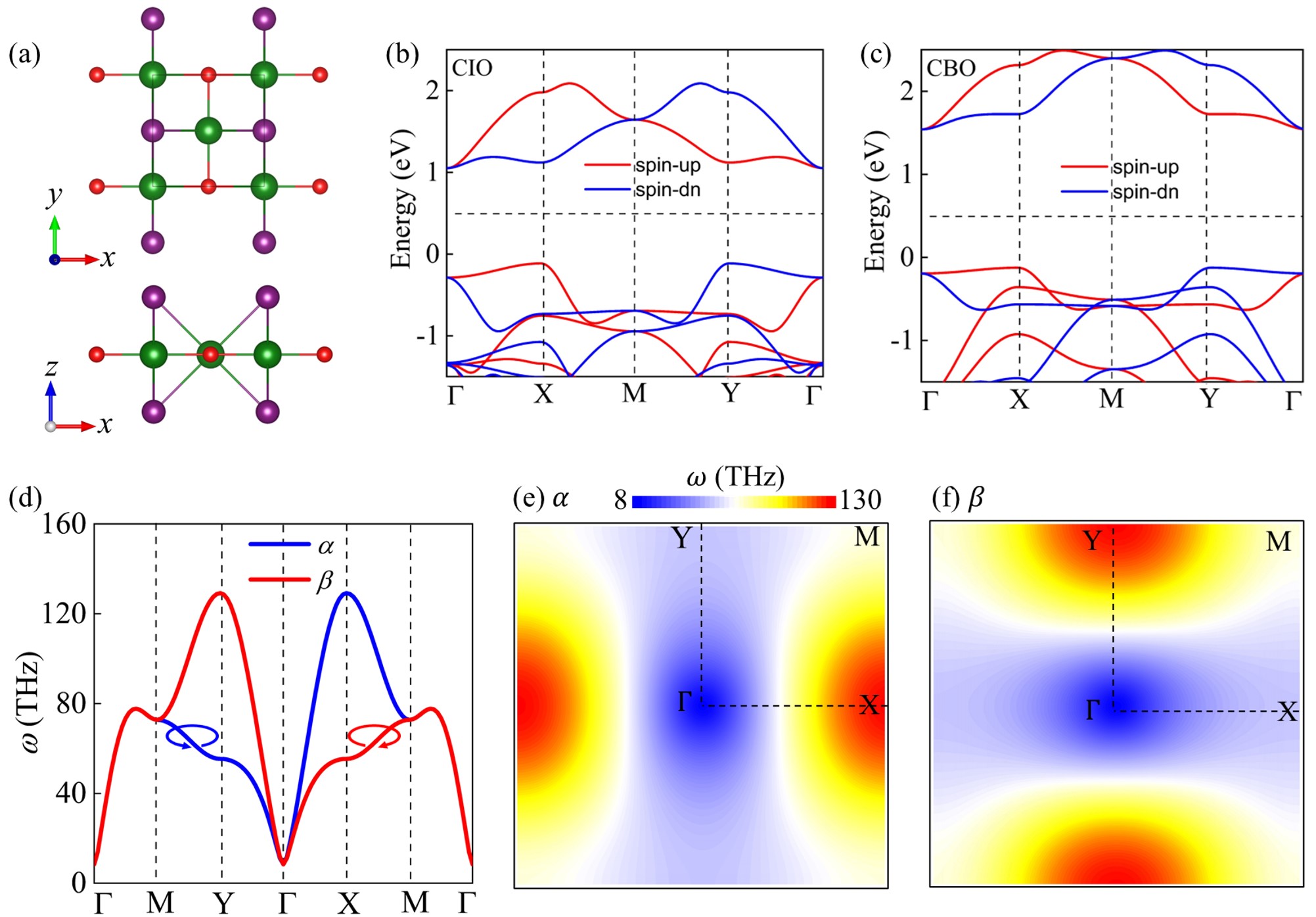}
\caption{(a) The top and side views of the crystal structure of $\rm Cr_2I_2O$. The green, purple, and red balls represent the Cr, I, and O elements, respectively. (b) and (c) The spin-polarized band structures. The band gap of $\rm Cr_2I_2O$ and $\rm Cr_2Br_2O$ reaches 1.24 and 1.85~eV, respectively. (d) Spin-splitting magnon band structure and (e, f) magnon energy spectra of $\rm Cr_2I_2O$.}
\label{FigS5}
\end{figure}

\textcolor{red}{Besides $\rm Cr_2Te_2O$ and $\rm Cr_2Se_2O$, we find that $\rm Cr_2I_2O$ and $\rm Cr_2Br_2O$ [see crystal structure in Fig.~\ref{FigS5}(a)] also combines non-relativistic spin splitting and insulating features in electronic band structures [Fig.~\ref{FigS5}(b) and (c)]. By using energy mapping methods, we obtain $J_1=12.25$, $J_2 =-1.91$, $J_2^{\prime}=6.20$ and $K=-0.07$~meV for $\rm Cr_2I_2O$, and $J_1=12.03$, $J_2 =-2.35$, $J_2^{\prime}=6.31$ and $K=0.02$~meV for $\rm Cr_2Br_2O$. Despite the anisotropic nearest-neighboring exchange coupling that exists in both candidates, only $\rm Cr_2I_2O$ preserves the out-of-plane magnetization. The anisotropic spin-momentum locking [Fig.~\ref{FigS5}(d-f)] leads to the spin Seebeck and spin Nernst effects independent of the external magnetic field and spin-orbit coupling. For $T=100$~K and $\tau_{0}=10$~ps, the maximum magnitude of $\sigma_{\parallel(\perp)}$ reaches1.64~meV/K.}

